\documentclass[preprintnumbers,amsmath,amssymb,aps]{revtex4}
\def\unit{\leavevmode\hbox{\small1\kern-3.6pt\normalsize1}}
 \normalsize

\def\lsim{\raise0.3ex\hbox{$\;<$\kern-0.75em\raise-1.1ex\hbox{$\sim\;$}}}
\def\gsim{\raise0.3ex\hbox{$\;>$\kern-0.75em\raise-1.1ex\hbox{$\sim\;$}}}

\usepackage{graphics}

\newcommand{\be}{\begin{eqnarray}}
\newcommand{\ee}{\end{eqnarray}}

\def\bea{\begin{eqnarray}}
\def\eea{\end{eqnarray}}

\usepackage{epsfig}
 \normalsize

\begin{document}

\title{Warm Dark Matter in B-L Inverse Seesaw}
\author{Amr El-Zant$^{1,2}$, Shaaban Khalil$^{1,3}$ and Arunansu Sil$^{4}$}
\affiliation{$^1$Center for Theoretical Physics, Zewail City for Science and Technology, 6 October City, Cairo, Egypt.\\
$^2$The British University in Egypt, Sherouk City, Cairo 11837, Egypt.\\
$^3$Department of Mathematics, Faculty of Science,  Ain Shams University, Cairo, Egypt.\\
$^4$ Department of Physics, Indian Institute of Technology,
Guwahati, Assam 781039, India.}
\date{\today}

\begin{abstract}
We show that a standard model gauge singlet fermion field, with
mass of order keV or larger, and involved in the inverse seesaw
mechanism of light neutrino mass generation, can be a good warm
dark matter candidate. Our framework is based on $B-L$ extension
of the Standard Model. The construction ensures the absence of
any mixing between active neutrinos and the aforementioned dark
matter field. This  circumvents the usual constraints on the mass
of warm dark matter imposed by X-ray observations. We show that
over-abundance of  thermally produced warm dark matter (which
nevertheless do not reach chemical equilibrium) can be reduced  to
an acceptable range in the presence of a moduli field decaying
into radiation --- though only when the reheat temperature is
low enough. Our warm dark matter candidate can also be produced
directly from the decay of the moduli field during
reheating. In this case, obtaining the right amount of relic
abundance, while keeping the
reheat temperature high enough as to be consistent with Big
Bang nucleosynthesis bounds, places constraints
on the branching ratio for the decay of the moduli field into dark matter.

\end{abstract}
\maketitle
\section{Introduction}

Recent Planck satellite observations of the fluctuations in
the cosmic microwave background~\cite{Ade:2013lta} suggest that
$26.8\%$ of the content of our Universe is in the form of dark
matter (DM). These observations, as well as measurements of large scale structure
 in the Universe, are consistent with the dark matter being made of relatively heavy (say $\sim 100$~GeV)
weakly interacting particles, which were in thermal equilibrium and
decoupled while non-relativistic. Indeed, the large scale distribution
of galaxies can be quite adequately explained by a model whereby 
the matter component of the Universe is dominated by such Cold
Dark Matter (CDM), with galaxies forming inside the potential well
of collapsed dark matter objects (halos), and mass largely following
the light traced by the galaxy distribution. This is a scenario that has developed over several
decades into a success story when it comes to explaining the gross structure of galaxies
and their large scale distribution~\cite{F-W}.

However, at smaller scales ($\lesssim 10$~kpc) the CDM-dominated structure formation scenario starts facing severe problems. Among these is
the dearth of observed dwarf galaxies, compared with the abundance of  satellite
halos inferred from CDM-based simulations~\cite{{Moore:1999nt},{Klypin:1999uc}}. Though solutions to this problem
have been proposed in the context of the astrophysics of galaxy
formation~\cite{{Bullock:2000wn},  {Benson:2001at},  {Somerville:2001km},   {Okamoto:2009rw},  {Wadepuhl:2010ib},   {Brooks:2012ah}},
a simpler and more natural step toward resolving this problem is to assume that
the dark matter is ${\it{warm}}$ \cite{wdm} instead of cold.
In this case, the free streaming length associated with less massive warm dark matter (WDM) particles at the era of
matter-radiation equality --- when structure formation becomes efficient --- is
generally larger, because such particles have higher velocities in kinetic
equilibrium. And within a distance of the order of the free streaming length~\cite{free-stream}, dark matter (DM)
particles can freely travel. Therefore, any structure formation at scales smaller than the
free streaming scale are naturally erased.

An accurate determination of the effective cutoff for  structure
formation at small scales  can be obtained from the scale length
at which the WDM affects the transfer function, filtering the
initial power spectrum of fluctuations \cite{{BBKS},
{Bode:2000gq}, {Viel:2005qj}, {Moore1}}. For a 100 GeV particle that was once in kinetic equilibrium 
and then decoupled,
the cutoff scale turns out to be too small to be relevant as far as galaxy
formation is concerned. A 1 keV WDM particle on the other hand
will not collapse into bound structures with mass below $10^{7}$
solar masses, and the mass function may be affected on mass scales
of two to three orders of magnitude larger~\cite{Moore1}. 
Substructure halos made of such particles also have smaller intrinsic
concentration, and therefore more in line with
observations~\cite{Lovell:2011rd}. Heavier WDM particles may not
solve this latter problem but, for masses up to a few tens of keV,
the mass function of small galaxy mass halos may still, in
principle, be affected. Mass scales smaller than keV on the other
hand are rather robustly ruled out by a combination of phase space
density constraints \cite{Shao:2012cg, boyarsk1, boyarsk2}
and Lyman-$\alpha$ observations \cite{Viel:2005qj,
Viel:2006kd, VielN}. In this context, a generous mass scale for
WDM that achieves kinetic equilibrium seems to be between 1-100 keV~\footnote{In this paper we generally
assume  that the dark matter particles achieve kinetic equilibrium, 
and so attain an equilibrium velocity distribution before decoupling 
(but not necessarily chemical equilibrium and associated equilibrium 
abundance). This constrains the WDM mass range. We discuss the case when 
kinetic equilibrium is not established  at the end of Section V.}.  
However many current model
candidates, e.g. sterile neutrinos, are further constricted
because the particles involved, mix with standard model particles.
This may give rise to significant X-ray emission, which then
constrains their mass below 3.5 keV \cite{XR1}. Indeed, barring the possibility 
that recent X-ray excesses detected  by the  XMM-Newton satellite  are due 
to  7~keV~ sterile neutrinos, these particles seem to be virtually ruled out, given the combination of
upper and lower mass constraints from X-ray and Lyman-$\alpha$
observations~\cite{VielN}.

In this paper we consider an alternative candidate for WDM that
may be naturally obtained within a simple extension of the
Standard Model (SM) gauge group with $U(1)_{B-L}$, where $B,L$
stand for baryon and lepton numbers
respectively~\cite{Marshak:1979fm,Mohapatra:1980qe,Khalil:2006yi}.
The evidence for non-vanishing neutrino masses, based on the
observation of neutrino oscillations \cite{oscillation}, indicates
that the SM requires an extension, as its left-handed neutrinos
are strictly massless due to the absence of right-handed neutrinos
and an exact global baryon minus lepton ($B-L$) number conservation.
The $U(1)_{B-L}$ extension of the Standard Model (SM) can generate
light neutrino masses through either Type-I seesaw or inverse
seesaw mechanism~\cite{Khalil:2006yi,Khalil:2010iu}. In the type-I
seesaw mechanism, right-handed neutrinos acquire Majorana masses at
the $B-L$ symmetry breaking scale, while in the inverse seesaw these
Majorana masses are not allowed by $B-L$ gauge symmetry;
another pair of SM gauge singlet fermions with tiny masses ($\sim
O(1)$ keV) must be introduced. One of these two singlet fermions
couples to the right handed neutrino and is involved in generating the
light neutrino masses. The other singlet is completely decoupled
and interacts only through the $B-L$ gauge boson and therefore can
serve the role of WDM.  Thus, the resulting WDM candidate,
being decoupled from the active neutrinos and all other SM
particles, is free from constraints such as X-rays bounds.

With  typical values of the annihilation cross section
associated with keV WDM one finds (within the standard cosmology)
that the thermal relic abundance (associated with chemical equilibrium) 
is quite high, as to be
inconsistent with cosmological observations. To circumvent this
problem, we consider the presence of a moduli field that decays
into radiation, with  WDM  being produced during reheating. We find
that, in order not to overproduce the relic abundance, a very low
reheat temperature ($\sim$ 3 MeV) is required for WDM particles
of masses of order 1 keV. For heavier candidates, reheat
temperatures must be tuned (unrealistically) below this limit in
order to produce acceptable relic abundances (unless one goes for
a much heavier DM having mass above 1 MeV).

The conclusions change however once we identify this WDM candidate
with the sterile neutrino in the inverse seesaw framework, since
the annihilation cross section involved is quite suppressed by the
heavy mass of the mediating gauge boson. We find that, for a  1~keV particle,  the desired relic abundance can be
naturally obtained with relatively large reheat temperature
($\sim$ 0.1 GeV), and that proper abundances can be
produced for the whole WDM mass range without violating Big Bang
nucleosynthesis constraints.  We also study the case where a scalar field can decay directly into WDM  (along with radiation).
It is shown that depending upon the branching ratios of the moduli
field decay into WDM, which must be rather modest if small enough
abundances are to result, the reheat temperature can vary
between 3 MeV to 100 MeV.

The paper is organized as follows: In the following section we
discuss the standard WDM production scenario and the associated abundance problem. Section
III is devoted to the calculation of WDM relic density in the
presence of a heavy field which decays into radiation. We discuss
the $B-L$ extension of the SM in the context of inverse seesaw and
identify the possible candidate for WDM coming out from the
construction itself in section IV. In section V, we estimate the
relic density of that WDM candidate in cases ranging pure thermal
production to predominantly nonthermal production, due to the decay 
of the scalar field into WDM particles and briefly discussing the case when the resultant WDM 
cannot be assumed to have attained kinetic equilibrium.
Finally we give
our conclusions in section VI.

\section{WDM as a relic of standard scenario}

In this section we briefly point out that the WDM --- we generically
name the WDM field as $\chi$ here and later on we will
identify it with the SM gauge singlet fermion field in the
context of inverse seesaw ---  relic abundance cannot be in
line  with the cosmological observations if it was in thermal
equilibrium.  The number density of WDM particles, $n_{\chi}$,
which were once in thermal equilibrium in the early Universe,
and decoupled when they were semi-relativistic or relativistic,
can be found by solving the following  Boltzmann equation:%
\be%
\frac{d n_\chi}{d t} + 3 H n_{\chi} = - \langle \sigma_\chi
v \rangle \left[n_\chi^2 - (n_\chi^{eq})^2\right],%
\label{Bol0}
\ee%
where $H$ is the Hubble  parameter and
$\langle \sigma_\chi v \rangle$ is the thermal average of
the annihilation cross section of the $\chi$ field multiplied by
the relative velocity of the two $\chi$ particles; $n_{\chi}^{eq}$
corresponds to the equilibrium value of $n_{\chi}$. Note that
the thermal equilibrium is preserved until the point where the
interaction rate $\Gamma(x) = n_{\chi} \langle \sigma_\chi
v\rangle$ ceases to be larger than $H$ (where the so
called decoupling happens). Defining the dimensionless
variable $x = \frac{m_{\chi}}{T}$, where $T$ represents the temperature, in a  radiation dominated universe with $g_*$ relativistic degrees of freedom,
$H(x)= \frac{\pi m^2_{\chi}}{x^2 M_{P}} \sqrt{\frac{g_*}{90}}$, where $M_P$ is 
the reduced Planck scale ($M_{P} = 2.4 \times 10^{18}$ GeV).
Being a few keV in mass, the particle under consideration may therefore decouple either relativistically ($x_F \gg O(1)$)
or semirelativistically ($x_F \sim O(1)$). The temperature at which $\chi$
decouples is the freeze out temperature $T_F$. As standard, we define
the ratio of the number density to entropy by $Y$. Since the relic abundance does not change much after decoupling (for a relativistically decoupled relic),
the final abundance is given by the equilibrium value, 
$Y_{\chi}$,
and hence
\be%
Y_{\chi, \infty} \equiv Y_{\chi,eq}(x_F).%
\ee%
The relic density is estimated as $\Omega_{\chi} = \frac{m_{\chi}
s_0 Y_{\chi, \infty }}{\rho_c}$, where $s_0$ and $\rho_c$ are the present entropy density and the critical density of the Universe.
It turns out that for $m_{\chi}$ in the region of our interest (1-100 keV), $\Omega_{\chi}
h^2 \gg 1$, which is inconsistent with observations; e.g., WMAP~\cite{wmap} and PLANCK~\cite{Ade:2013lta} results
suggest $\Omega h^2 \simeq 0.12$.

If  the WDM decouples semi-relativistically, {\it i.e.,}
$T_F \sim m_\chi$, the final relic abundance would depend on the freeze
out temperature. To determine that, we need to know the form
of the  cross section involved. In \cite{drees}, it is shown that in this case,
a thermally averaged annihilation cross section can be approximated, over a large range in temperature,  as %
\be%
\langle \sigma_\chi v \rangle \simeq \frac{G_F^2 m_\chi^2}{16
\pi}\
\left (\frac{12}{x^2} + \frac{3 + 6x}{(1+x)^2} \right ),%
\label{cs-approx}%
\ee%
where $G_F$ is the Fermi coupling constant
($\sim 1.2 \times 10^{-5}$ GeV$^{-2}$ ) of four-fermion
interaction (between the WDMs and SM particles, mostly into light
neutrinos when $m_{\chi} \simeq $ few keV). Then it turns
out that $\langle \sigma_\chi v \rangle_F$ is
 of order $10^{-20}\ {\rm GeV}^{-2}$, which is
many order of magnitudes below the desired value, making the final
relic density $\Omega_{\chi} h^2 \gg 0.12$. In this context, one
can conclude that it is not possible to get the right amount
of relic abundance with WDM that was in thermal equilibrium.

In the next section we show that one may  get around the
problem of  large relic abundance by considering the
existence of a long-lived particle that dominates the universe prior to
its decay. For example, a scalar field $\phi$ (possibly an inflation
or a moduli field in supersymmetric models), oscillating around
its true minimum, would dominate the energy density of the
universe. We consider that the WDM is produced
during this reheating era; and  expect that the WDM relic density would be reduced due to
the entropy release from the decay of $\phi$. The decay of
the field depends on the coupling with the SM fields and other
particles. Once the $\phi$-field decays away completely, we are
left with radiation domination. 
\section{Nonequilibrium Production of WDM during reheating}
\label{sec:low}

Our universe may have gone through
one or more inflationary phases, which are followed by a reheating
stage, whereby a scalar field $\phi$ decays into radiation~\cite{{Bassett:2005xm},{GKR}}.
The reheat temperature can be related to the decay width ($\Gamma_\phi$)
of $\phi$   through %
\be T_{RH}= \left(\frac{90}{\pi^2 g_*(T_{RH})}\right)^{1/4}
(\Gamma_\phi M_{P})^{1/2}\ . \label{trh} %
\ee
We consider the case where our WDM candidate $\chi$ is produced during
reheating. To obtain the associated abundance,  one has to solve a system 
of Boltzmann equations for
density of the WDM ($\chi$),  the $\phi$-field and the radiation ($R$), which 
(assuming kinetic equilibrium) are given by  \cite{{GKR},{Gelmini:2006pw},{Moroi:1999zb},{Chung:1998rq}}: %
\bea
\dot{\rho_{\phi}} & = & -3H\rho_{\phi} - \Gamma_{\phi} \rho_{\phi}, \\
\dot{\rho_{R}} & = & -4H\rho_{R} + \Gamma_{\phi} \rho_{\phi} + \langle \sigma_{\chi} v \rangle 2 \langle E_{\chi} \rangle 
\left[ (n_{\chi})^2 -(n_{\chi}^{eq})^2 \right], 
\label{eq6}\\
\dot{n_{\chi}} & = & -3Hn_{\chi} - \langle \sigma_{\chi} v \rangle
\left[ (n_{\chi})^2 -(n_{\chi}^{eq})^2 \right],%
\eea %
where the time derivative is denoted by the dot and $\langle E_{\chi}\rangle$ 
is the average energy associated with each $\chi$, given by the expression 
$\simeq \sqrt{{m_{\chi}^2} + 9 T^2}$. Here $\rho_{R}$ and
$\rho_{\phi}$ represent the energy densities of
radiation component and the moduli field $\phi$ respectively. Since our WDM candidate $\chi$ is a stable particle,
we do not need to consider its decay. 

Following~\cite{GKR}, we introduce the normalized variables involving the scale factor of the Universe ($a$), such that
\be%
\Phi \equiv \rho_{\phi} T^{-1}_{RH} a^3; R \equiv \rho_R a^4; X
\equiv n_{\chi} a^3; A \equiv a/a_I. %
\ee %
The label $I$ corresponds to the initial condition and $a_I$ is chosen to be
$T^{-1}_{RH}$ for convenience. In terms of these variables, the above
set of Boltzmann equations becomes, %
\bea %
\frac{d\Phi}{dA} & = & -\left
( \frac{\pi^2 g_*}{30}\right )^{1/2} \frac{A^{1/2}
\Phi}{\sqrt{\Phi + R/A + X \langle E_{\chi}\rangle/T_{RH}}}, \\
\frac{dR}{dA} & = & \left ( \frac{\pi^2 g_*}{30}\right )^{1/2}
\frac{A^{3/2} \Phi}{\sqrt{\Phi + R/A + X \langle E_{\chi}\rangle/T_{RH}}} + 
\frac{{\sqrt{3}}M_P A^{-3/2}\langle \sigma_{\chi} v \rangle 2\langle E_{\chi} \rangle \left[
X^2 -(X_{eq})^2 \right] }{\sqrt{\Phi + R/A + X \langle E_{\chi}\rangle/T_{RH}}}, \\
\frac{dX}{dA} & = & - 3^{1/2}
\frac{A^{-5/2} \langle \sigma_{\chi} v \rangle M_{P}
T_{RH}}{\sqrt{\Phi + R/A + X \langle E_{\chi}\rangle/T_{RH}}} \left[
X^2 -(X_{eq})^2 \right].%
\label{Boltz}
\eea%
The Hubble parameter is $H = \left ( \rho_{\phi} + \rho_{R}+ \rho_{\chi} \right)^{1/2}/\left({\sqrt{3} M_P}\right)$, 
where we assume that $\rho_\chi = \langle E_{\chi}\rangle n_{\chi}$. 

Before the onset of $\phi$ decay (characterized by time $H^{-1}_I$), the
energy density of the universe is completely dominated by the $\phi$ field and thus
the following initial conditions can naturally be adopted\footnote{We generally
started with $R=0$ as initial condition, except when this caused numerical
problems, in which case $R$ was started with
a very small value of the order of machine precision. The
results were verified as independent of the particular value
chosen.}: \\
\be 
\Phi_I = \frac{3 M^2_P H^2_I}{T^4_{RH}}, R_I =
X_I = 0, A_I =1.
\ee

In the  period between $H^{-1}_I$ and $\Gamma^{-1}_{\phi}$ (indicating the completion of the $\phi$ decay)
the Hubble expansion parameter $H$ can be written as~\cite{GKR}
\be H = \sqrt{\frac{5 \pi^2
g_*^2(T)} {72 g_*(T_{RH})}} \frac{ T^4} {T_{\rm
RH}^2 M_{P}}, %
\label{H-defn}
\ee%
from which we can obtain the $H_I$ by replacing $T$ by $T_{\rm max}$. Here $T_{\rm \max}$ is the maximum temperature achieved during
reheating (it is generally greater than  $T_{RH}$). The results are quite insensitive to the choice of $T_{\rm max}$, as long as $T_{\rm max} \gg m_{\chi}$. In our calculations, we assume $T_{\rm max} = 10$~GeV. The temperature is inferred from the relation %
\be %
T = \left(\frac{30}{ \pi^2 g_*(T)} \right)^{1/4} \frac{R^{1/4}}{A}
T_{RH}.%
\ee%

\begin{figure}[t]
\begin{center}
\epsfig{file=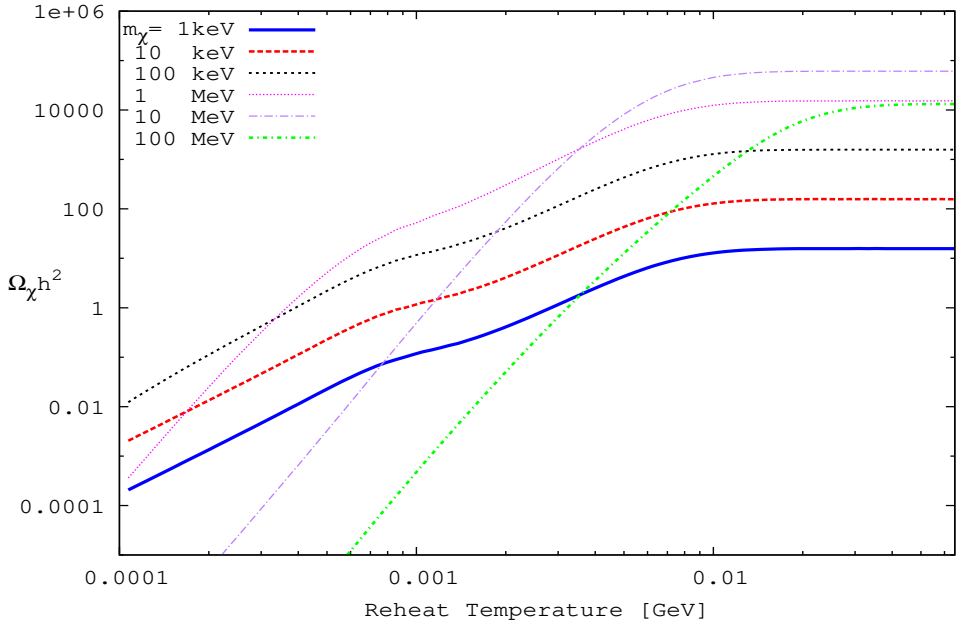,height=8.5cm,width=13.5cm,angle=0}
\caption{Relic abundance of the dark matter ($\chi$) as function of the reheat temperature
for different values of $m_{\chi}$, in the presence of a decaying field $\phi$. The region 
below the reheat temperature 1 MeV is not of physical importance (kept just for the 
demonstrative purpose to show the variation), as it would be inconsistent with the Big Bang 
nucleosynthesis bound.}
\label{Fermi}
\end{center}
\end{figure}

The behavior of $X$ can be roughly understood by considering Eq.~(\ref{Boltz}), while the $\phi$-field is dominant
(during reheating). 
Assuming that $\chi$ particles do not reach thermal equilibrium\footnote{We will discuss this again
in Section V, Fig.5.} before the reheat temperature $T_{RH}$ is reached~(${\it {i.e.}} X \ll X_{eq}$),  Eq.~(\ref{Boltz}) can be written as
\be
\frac{dX}{dA} = C A^{-5/2} \langle \sigma_{\chi} v \rangle X_{eq}^2,
\label{ab:red}
\ee
where $C = \sqrt{\frac{3}{\Phi_I}} M_{P} T_{RH}$. Note that in this case $X$ (and hence the final relic abundance $\Omega_{\chi} h^2$)  is proportional to
$\langle \sigma_{\chi} v \rangle$ instead of the standard inverse
dependence. The dependence on reheat temperature is also important to notice.

The results are depicted in Fig.~\ref{Fermi}, where the final contribution
of dark matter particles to the mass density as a function of reheat temperature is plotted for different masses of the DM particle.
These were obtained  numerically (keeping all the terms in the set of Boltzmann equations)
where the cross section is characterized by standard Fermi coupling (with cross section largely following  Eq.~(\ref{cs-approx}); cf. Fig.~3).
The number of relativistic degrees of freedom is approximated as a
step function of temperature. We start  with $g_* = 2$ and  when
the temperature increases (or eventually decreases), as the
calculation proceeds, $g_*$ is accordingly modified. Due to the
discontinuity, the calculation is stopped and started with the new
value of $T$, keeping other parameters constant. For most of the parameter range discussed here the equations are
stiff and require an implicit scheme; we use a
backward differentiation method (e.g., \cite{NR}), with relative
tolerance $10^{-8}$ per time step. The integration is stopped once
the comoving density $X$ converges to a constant asymptotic value
with the same relative precision as the integration itself,
provided reheating is complete (that is when $\phi$ becomes
vanishingly small) and all variations in $g_*$ have ceased. In
this case the present number density of particles is given by
\be
n_{\chi} = \frac{X}{a_c^3}~\frac{a_c^3}{a_o^3} =
\frac{X}{a_c^3}~\frac{T_0^3}{T_c^3},
\ee
where $T_0$ is the
current CMB temperature, $T_c$ is the temperature at $X$
convergence (during the radiation era) and $a_c$ is the
corresponding scale factor,  given by $a_c = a_I A_c = A_c/T_{RH}$.
The mass density is $\rho_{\chi} = m_{\chi}  n_{\chi}$ and the
contribution to the critical density is $\Omega_{\chi} =
\rho_{\chi}/\rho_{\rm crit}$, with $\rho_{\rm crit} = 3 H_0^2/ 8
\pi G$, so that  $\Omega_{\chi} h^2 = \frac{8\pi G \rho_{\chi}}{3}
(\frac{h}{H_0})^2$. Here the scaled Hubble parameter, $h$, is defined 
through the Hubble constant at the present epoch, $H_0$, by $H_0 = 100 h$ 
km $s^{-1}$ Mpc$^{-1}$.  

In the cases considered in this section the DM particle production mechanism is thermal, since the $\phi$-field does not decay into WDM. 
However that does not necessarily imply that the particle has to reach chemical equilibrium, as will become clear from the following discussion.
We divide the discussion into two  parts. \noindent {\bf{I.}} Region  through which 1~ ${\rm{keV}} \le
m_{\chi} <  1~ {\rm{MeV}}$: First note that the physically relevant region in this case  corresponds
to $m_{\chi} < T_{RH}$ when $T_{RH} > 1$ MeV (to satisfy the 
Big Bang nucleosynthesis bounds). The curves in Fig. {\ref{Fermi}} can then be explained following
the analytical result obtained in \cite{GKR}. As found in
\cite{GKR}, $\Omega_{\chi} h^2$ is proportional to
$T_{\rm RH}^3/ g^{3/2}_* (T_{\rm RH})$ once the thermally averaged
cross-section in Eq.~(\ref{cs-approx}) is considered, with $g_*
(T_{\rm RH})$ referring to the number of relativistic degrees of
freedom at reheating. Deviations from pure power law in our
numerical calculations reflect the detailed evolution of $g_*$
as the full equations are integrated for different reheat temperatures. The rising sections of the curves
correspond to the production of relativistic particles that do not reach chemical equilibrium. The dependence of $\Omega_{\chi} h^2$ on $T_{RH}$
flattens for higher range of $T_{RH}$. This is due to the fact
that, as the reheat temperature is increased the DM particles
come closer and closer to attaining chemical equilibrium.
In this regime of relatively large $T_{RH}$, one recovers the
abundance estimates associated with the standard scenario, which
are  again seen to be unrealistically large unless the particle mass is
reduced to the eV, or hot dark matter, range. The inferred abundances for WDM particle mass in the keV range
can indeed be consistent with observations only for  $T_{RH}$ so low as to be 
marginally compatible with constraints from Big Bang
nucleosynthesis.

\noindent {\bf{II.}} Region through which $1~ {\rm{MeV}} \le
m_{\chi} \le  100~ {\rm {MeV}}$: In this case $m_{\chi}$ could either
be larger or smaller than $T_{RH}$.  For large particle masses
relative to $T_{\rm RH}$, the bulk of particle production occurs
while the DM particle in consideration is non-relativistic and the
inferred densities are proportional to $T_{\rm RH}^7$ \cite{GKR}.
The curve again flattens off as chemical equilibrium is reached
after reheating. The case of $m_\chi = 1~{\rm MeV}$ is
intermediate; particle  production is predominately in the
relativistic regime for large $T_{\rm RH}$ and non-relativistic
for smaller $T_{RH}$, with a transition around $T_{\rm RH} \sim
m_\chi = 1~ {\rm MeV}$. The asymptotic densities are in line with
estimates inferred in the context of the standard scenario using
the same form of the cross-section employed in \cite{drees}, and
are unrealistically large. Thus, from Fig.~\ref{Fermi}, we
conclude that the DM relic densities are only compatible with
observations for very low reheat temperatures which is barely
consistent with Big Bang nucleosynthesis. Furthermore, this only
happens when the particle mass is of order 10~MeV or larger.

 Thus, in the context considered here, particle masses are constrained either in the keV range or below and for reheat
temperatures marginally consistent with nucleosynthesis constraints, and  in
possible tension with Lyman-$\alpha$ constraints \cite{Viel:2006kd, VielN},  or in the range of 10~MeV or above,
in which case they would not ameliorate problems connected with
galaxy formation.  Below,  however, we will replace this Fermi coupling
by a $B-L$ one. In that case, since the mediator would be a sufficiently heavy particle (the gauge boson of $B-L$), the corresponding
cross section, $\langle \sigma v \rangle$, of $\chi$ will be suppressed.
This would further suppress the relic density as it is proportional
to $\langle \sigma v \rangle$ for the non-equilibrium production of $\chi$ during reheating. So in that case, we expect $\Omega_{\chi} h^2$ 
to enter within the allowed range by WMAP and PLANCK. 
\noindent

\section{WDM in $B-L$ Extension of the SM with Inverse Seesaw}

As advocated in the introduction, one of the most attractive
mechanisms that can naturally accommodate small neutrino masses
with TeV scale right-handed neutrinos is what is known as the ``Inverse
Seesaw Mechanism". This class of models predicts the following two
types of neutrinos, besides the SM-like light neutrinos: $(i)$
Heavy (TeV) neutrinos, which are quite accessible and have
interesting phenomenological implications; $(ii)$ Sterile light
(keV) neutrinos, which has zero mixing with active neutrinos.

In this section we first show how the inverse seesaw mechanism can
be naturally embedded in a low scale gauged $U(1)_{B-L}$ extension
of the standard model. We also argue that it provides a natural
candidate for WDM. The $B-L$ extension of the SM is based on the
gauge group: $SU(3)_C\times SU(2)_L\times U(1)_Y\times U(1)_{B-L}$
\cite{Khalil:2006yi,b-l,Khalil:2010iu}. The standard model is
characterized by global $U(1)_{B-L}$ symmetry. If this symmetry is
locally gauged, then the existence of three SM singlet fermions
(the right-handed neutrinos) is a quite natural assumption to make
in order to cancel the associated anomaly, which is a necessary
condition for the consistency of the model. The extra $U(1)_{B-L}$
is spontaneously broken by a SM singlet scalar $\eta$ with $B-L$
charge $=-1$ \cite{Khalil:2010iu}. The model therefore naturally
predicts one extra neutral gauge boson $Z'$ corresponding to $B -
L$ gauge symmetry. In addition, three SM singlet fermions
$\nu_{R_i}$ with $B-L$ charge $=-1$ are introduced for the
consistency of the model. Finally, three SM pairs of singlet
fermions $S_{1,2}$ with $B-L$ charge $=\mp2$, respectively, are
introduced to implement the inverse seesaw mechanism
\cite{Khalil:2010iu}. The $B-L$ quantum numbers of fermions and
Higgs bosons of this model are summarized in Table~\ref{tab:b-l}. 
A discrete symmetry $Z_4$ is also considered in order to forbid several 
unwanted terms. The $Z_4$ charge assignment is included in 
Table ~\ref{tab:b-l}.
\begin{table}[thbp]
\centering
\begin{tabular}{||c|c|c|c|c|c|c|c|c|c|c|c||}
\hline\hline ~~Particle~~ & ~~$Q$~~ & ~~$u_R$~~& ~~$d_R$~~& ~~$L$~~ & ~~$e_{R}$~~ & ~~$\nu _{R}$ ~~& ~~$h$~~ & ~~$\eta$~~ & ~~$S_1$~~ & ~~$S_2$ ~~\\
\hline
$Y_{B-L}$& $1/3$  & $1/3$ & $1/3$ & $-1$ & $-1$ & $-1$ & $0$ & $-1$ & $-2$ & $+2$\\
\hline
$Z_{4}$& $1$  & $1$ & $1$ & $-i$ & $-i$ & $-i$ & $1$ & $i$ & $-1$ & $1$\\
\hline\hline
\end{tabular}%
\caption{$B-L$ quantum numbers and $Z_4$ charges of fermions and Higgs particles.}
\label{tab:b-l}
\end{table}
The Lagrangian of the leptonic sector in this model is given by
\cite{Khalil:2010iu}
\begin{eqnarray}
{\cal L}_{B-L}&=&-\frac{1}{4} F'_{\mu\nu}F'^{\mu\nu} + i~
\bar{L} D_{\mu} \gamma^{\mu} L + i~ \bar{e}_R D_{\mu}
\gamma^{\mu} e_R + i~
\bar{\nu}_R D_{\mu} \gamma^{\mu} \nu_R\nonumber\\
&+& i~ \bar{S}_{1} D_{\mu} \gamma^{\mu} S_{1} + i~ \bar{S}_{2}
D_{\mu} \gamma^{\mu} S_{2}
+(D^{\mu} h)^\dagger D_{\mu} h + (D^{\mu} \eta)^\dagger D_{\mu}\eta -V(h, \eta)\nonumber\\
&-&\Big(\lambda_e \bar{L} h\, e_R+\lambda_{\nu}
\bar{L} \tilde{h} \nu_R +\lambda_{S} \bar{\nu}^c_R \eta
S_2 \Big) + h.c.,
\label{lagranigan}
\end{eqnarray}
where $F'_{\mu\nu}$ is the field strength of the $U(1)_{B-L}$ and $h$ is the SM higgs field.
{Note that the $B - L$ symmetry allows a mixed kinetic term $F_{\mu\nu} F'^{\mu\nu}$. 
This term leads to a mixing between $Z$ and $Z'$. However due to the stringent
constraint from LEP II on $Z-Z'$ mixing \cite{Carena:2004xs}, one may neglect this term \cite{khalil-1105}. Therefore, after the 
$B-L$ and the EW symmetry breaking, through
non-vanishing vacuum expectation values (VEVs) of $\eta$:
$|\langle\eta\rangle|= v'/\sqrt 2$ and $h$:
$|\langle h \rangle|= v/\sqrt 2$, one finds that the neutrino
Yukawa interaction terms lead to the following mass
terms \cite{Khalil:2010iu}:%
\be%
{\cal L}_m^{\nu} = m_D \bar{\nu}_L \nu_R + M_N \bar{\nu}^c_R S_2 + h.c.,%
\ee%
where $m_D=\frac{1}{\sqrt{2}}\lambda_\nu v$ and $ M_N =
\frac{1}{\sqrt 2}\lambda_{S} v' $. Here $v'$ is assumed to be of
order TeV, which is consistent with the result of radiative $B-L$
symmetry breaking found in gauged $B-L$ model with supersymmetry
\cite{Khalil:2007dr} and $v =246$ GeV. Note that the spontaneous
$B-L$ symmetry breaking leads to the following $Z'$ gauge boson
mass $M^2_{Z'}=4g^2_{B-L} v^{\prime 2}$  ~($g_{B-L}$ is the gauge
coupling of $U(1)_{B-L}$). The experimental search for $Z'$ LEP II
\cite{Carena:2004xs} implies that $ M_{Z'}/g_{B-L}>6\ {\rm TeV}$.

In addition one may generate very small Majorana masses for
$S_{1,2}$ fermions through possible non-renormalizable terms like
$\bar{S}^c_{1} {\eta^\dag}^{4} S_{1}/M^3$ and $\bar{S}^c_{2}
{\eta}^{4}
S_{2}/M^3$. Note that the smallness of these masses are ensured by the choice 
of $Z_4$ charges of the fields involved. Therefore the Lagrangian of neutrino masses, in the flavor basis, is given by %
\be%
{\cal L}_m^{\nu} =\mu_s \bar{S}^c_2 S_2 +(m_D \bar{\nu}_L \nu_R + M_N \bar{\nu}^c_R S_2 +h.c.) ,%
\ee%
where $\mu_s=\frac{v'^4}{4 M^3}\lsim 10^{-6}$ GeV. The choice of the discrete symmetry 
also forbids a possible large mixing term $m S_1 S_2$ in the Lagrangian which could 
otherwise spoil the inverse seesaw structure.
Therefore, the
neutrino mass matrix can be written as ${\cal M}_{\nu}
\bar{\psi}^c \psi$
with $\psi=(\nu_L^c ,\nu_R, S_2)$ where ${\cal M}_{\nu}$ is approximately 
given by, %
\be {\cal M}_{\nu} \simeq
\left(%
\begin{array}{ccc}
  0 & m_D & 0\\
  m^T_D & 0 & M_N \\
  0 & M^T_N & \mu_s\\
\end{array}%
\right). %
\label{inverse}
\ee%
A few additional non-renormalizable terms can also 
be present. However, being small, their contributions are not incorporated in $M_{\nu}$ above. For example, 
terms like $LLhh{\eta^{\dagger}}^2 /M^{3}$ and  $\bar{L}hS_2{\eta}^3 /M^{3}$ can contribute to 
11 and 13 (or 31) entries of $M_{\nu}$ respectively. 
Nevertheless, the contribution of $LLhh{\eta^{\dagger}}^2 /M^{3}$ is 
proportional to $\left(v/v'\right)^2\mu_s$ and that of the other term is  
proportional to $\left(v/v'\right)\mu_s$. Since $\left(v/v'\right) \lesssim \cal{O}$ $\left(10^{-1}\right)$ 
and $\mu_s \sim 10^{-6}$ GeV only, 
these terms do not have much impact on the overall structure of $M_{\nu}$. 
Also there would be a small contribution to the right handed neutrino Majorana mass term, $m_R$,
in the 22 entry of $M_{\nu}$, originating from ${\bar {\nu_R^c}} \nu_R {\eta^{\dagger}}^2/M $
that can not be prevented with the use of $Z_4$. However $m_R$ being small compared to
the $m_D$ and $M_N$, its presence will not alter the light neutrino mass eigenvalues obtained from
the above structure (Eq.(\ref{inverse})) to the leading order \cite{law}.

The diagonalization of the mass matrix in Eq.(\ref{inverse}) with nonzero 
$m_R$ leads \footnote{We have checked numerically that the presence of 
$m_R$ (which is larger than $\mu_s$ however smaller compared to $m_D$ and $M_N$) 
does not alter the form of Eq.(\ref{mnul}). Similarly the nonzero 11, 13, 31 entries 
(which are suppressed by 1$/M^3$) are numerically insignificant to produce 
any impact on the light neutrino mass.} to the following light and heavy 
neutrino masses \cite{Khalil:2010iu} respectively in the leading order, 
under the consideration $m_R, \mu_s \ll m_D, M_N$ ~ \cite{law}: %
\begin{eqnarray}%
m_{\nu_l} & \simeq & m_D M_N^{-1} \mu_s (M_N^T)^{-1} m_D^T,\label{mnul}\\
m_{\nu_H}^2 & = & m_{\nu_{H'}}^2 \simeq M_N^2 + m_D^2. %
\end{eqnarray} %

\begin{figure}[t]
\begin{center}
\epsfig{file=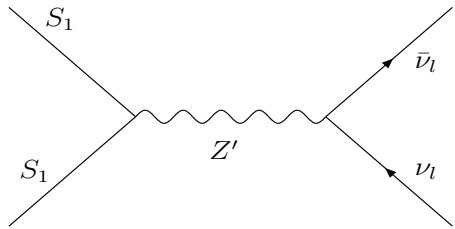,height=3.cm,width=6.cm,angle=0}
\label{sigma}%
\caption{Feynmann diagram for $S_1$ annihilation into light
neutrinos, vis $Z'$ gauge boson.} \label{S1annihilation}
\end{center}
\end{figure}

\noindent On the other hand, the second SM-singlet fermion, $S_1$, remains
light with mass given by%
\be%
m_{S_1} = \mu_s \simeq {\cal O}(1)~ {\rm keV}.%
\ee%
It is important to note that the $S_1$ is a kind of sterile neutrino
that has no mixing with active neutrinos. It only interacts
with the gauge boson $Z'$. Therefore, it is free from all
constraints imposed on sterile neutrinos due to their mixing with
the active neutrinos \cite{boyarsk2}. These constraints come
mainly from the one loop decay channel into photon and active
neutrino, which would produce a narrow line in the
diffuse $\gamma$ and $X$ rays background radiation. This in turn implies
that the mixing angle between sterile and active neutrinos is
limited by \cite{boyarsk2}
$$\theta \lsim 1.8 \times 10^{-5} \left(\frac{\rm keV}{M_{\rm
sterile}}\right)^5.$$ In contrast, the masses of our sterile
neutrinos ($M_{\rm sterile}$) are not restricted to the keV range. Being
odd under the $Z_2$, $S_1$ is a stable particle. So it is a quite natural candidate for
warm dark matter. It annihilates through one
channel only, into two light neutrinos mediated by $Z'$, as shown
in Fig. \ref{S1annihilation}. The thermal averaged annihilation
cross section of $S_1 S_1 \to
\nu_l \nu_l$ is given by \cite{gelmini-gondolo-1009}%
\be%
 \langle\sigma_{S_1}^{ann} v\rangle = \frac{\int_0^\infty dp~ p^2
W_{S_1S_1}(s) K_1(\frac{\sqrt{s}}{T})}{m_{S_1}^4 T \left[K_2(\frac{m_{S_1}}{T})\right]^2} , %
\label{S1sigma}\ee%
where $W_{S_1S_1}$ is defined as the annihilation rate per unit
volume and unit time \cite{seto} through %
\be %
W_{S_1S_1}(s) = \frac{1}{32 \pi} \int \frac{d \cos \theta}{2}
\sqrt{\frac{s-4 m_{S_1}^2}{s}} |{\cal M}(S_1 S_1 \to \nu_l
\nu_l)|^2,%
\ee%
with
\be%
\int \frac{d \cos\theta}{2}|{\cal M}(S_1 S_1 \to \nu_l \nu_l)|^2
 = \frac{2}{3} \frac{|g_{B-L}^2 q_{S_1} q_{\nu}|^2 }{(s-M_{Z'}^2)^2 + M_{Z'}^2
\Gamma_{Z'}^2} (s-4m_{S_1}^2)s . \label{sigma}%
\ee%
\begin{figure}[t]
\begin{center}
\epsfig{file=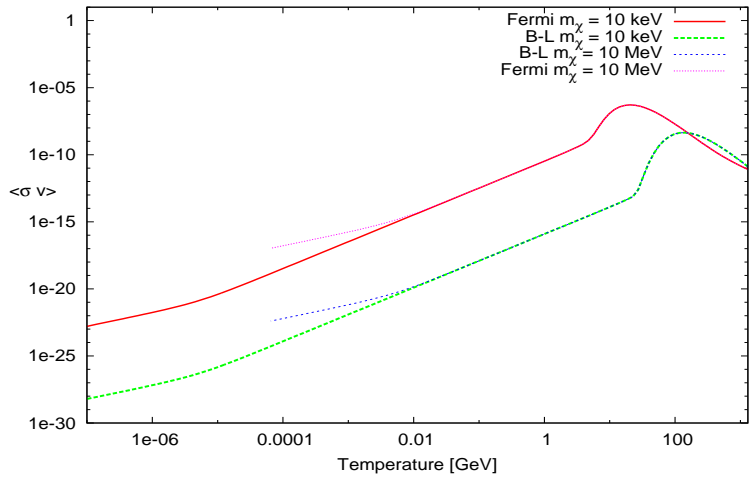,height=6.5cm,width=10.5cm,angle=0}
\caption{The thermally averaged annihilation cross section $\langle 
\sigma v \rangle$ (in GeV$^{-2}$) as function of the
temperature (in GeV) when the WDM particle under consideration interacts through  i) a Fermi coupling
or  ii) B-L coupling.} \label{figsigma}
\end{center}
\end{figure}
\noindent Here $s = 4(m^2_{S_1} + p^2)$ is the center of mass energy squared and $K_1, K_2$ are 
the modified Bessel functions. $q_{S_1}=2$, $q_{\nu}=-1$ are the $B-L$ charges. The gauge 
boson $Z'$ has the mass $M_{Z'}$, would be of order ${\cal
O}(100)$ GeV for $g_{B-L} = {\cal O}(0.1)$, and its decay width is 
$\Gamma_{Z'}$. In Fig. \ref{figsigma} we display the thermal averaged cross
section $\langle \sigma^{ann}_{S_1} v \rangle $ as function of
temperature $T$ for $m_{\chi} = 10$ keV and 10 MeV with $M_{Z'}= 600$
GeV and $g_{B-L} = 0.1$. In this plot, we also include the
corresponding cross-section for a similar candidate of DM, when it
has a standard Fermi coupling (with $M_Z = 91.2 ~{\rm GeV}$, $\Gamma_Z = 2.5 ~{\rm GeV}$ 
and $M_W = 80.4 ~{\rm GeV}$),  
for illustration. 
As can be seen from Fig. \ref{figsigma}, $\langle
\sigma^{ann} v \rangle$ for the $B-L$ model is  generally five to six orders of magnitude
smaller than the  cross section with Fermi coupling considered in the previous
section, which is expected as the suppression factor turns out to be 
$\sim \left( g_{B-L}/g_F \right)^4 \left( M_Z/M_{Z'}\right)^4 \sim 10^{-6}$ ~
($g_F$ is the gauge coupling constant in case of Fermi coupling). Note that 
this conclusion is largely independent of temperature and particle mass as well.

\section{WDM production during reheating in the $B-L$ model}

In this section we study WDM production in the context of the $B-L$ model
described above. In Section III it was impossible to consider the WDM
particles as direct products of the  $\phi$ field decay, without overproducing
the DM content of the Universe. 
In the present case, we will show that this constraint can be relaxed. This is primarily 
because the cross section $\langle \sigma^{ann} v \rangle$ is different;
we assign the WDM candidate with the $S_1$ field involved in
inverse seesaw, and the related cross-section is approximately
five or six orders of magnitude smaller compared to the case considered
in section III with Fermi coupling. So we naturally expect that
the dark matter abundances may radically differ from those
inferred in section III. The prediction involves whether the dark
matter particle reaches chemical equilibrium before reheating or
not, whether the main particle production happens when the
concerned particle is relativistic or not, and whether the field
directly decays into the dark matter particle. We have tried to
address these points in the rest of this section.

In order to realize WDM production through direct decay of the $\phi$-field, one should assume that the field has a strongly
suppressed coupling with the WDM, $\chi$. Then $\Gamma_{\phi}$
describes the total decay width of $\phi$, inclusive of the decay
into DM. We thus define the branching ratio of
the $\phi$ field by $B_{\chi}$. The Boltzmann
equations considered in section III are now replaced by the corresponding set
\cite{Khalil:2002mu},

\bea
\dot{\rho_{\phi}} & = & -3H\rho_{\phi} - \Gamma_{\phi} \rho_{\phi}, \\
\dot{\rho_{R}} & = & -4H\rho_{R} + (1 - B_{\chi}) \Gamma_{\phi} \rho_{\phi} +\langle \sigma^{ann}_{\chi} v \rangle
2\langle E_{\chi} \rangle \left[ (n_{\chi})^2 -(n_{\chi}^{eq})^2 \right], \\
\dot{n_{\chi}} & = & -3Hn_{\chi} + \frac{B_{\chi}}{m_{\chi}}
\Gamma_{\phi} \rho_{\phi} - \langle \sigma^{ann}_{\chi} v \rangle
\left[ (n_{\chi})^2 -(n_{\chi}^{eq})^2 \right], \label{eqchi}%
\eea %
where $(1 - B_{\chi})$ represents the decay into radiation. In terms of
the previously defined normalized variables, the above set of equations
can now be written as,
\bea %
\frac{d\Phi}{dA} & = & -\left
( \frac{\pi^2 g_*}{30}\right )^{1/2} \frac{A^{1/2}
\Phi}{\sqrt{\Phi + R/A + X \langle E_{\chi}\rangle/T_{RH}}}, \label{Boltz1} \\
\frac{dR}{dA} & = & \left ( \frac{\pi^2 g_*}{30}\right )^{1/2}
\frac{\left( 1-B_{\chi}\right) A^{3/2} \Phi}{\sqrt{\Phi + R/A + X \langle E_{\chi}\rangle/T_{RH}}} + 
\frac{{\sqrt{3}}M_P A^{-3/2}\langle \sigma^{ann}_{\chi} v \rangle 2\langle E_{\chi} \rangle \left[
X^2 -(X_{eq})^2 \right] }{\sqrt{\Phi + R/A + X \langle E_{\chi}\rangle/T_{RH}}}, \\
\frac{dX}{dA} & = & - 3^{1/2}
\frac{A^{-5/2} \langle \sigma_{\chi} v \rangle M_{P}
T_{RH}}{\sqrt{\Phi + R/A + X \langle E_{\chi}\rangle/T_{RH}}} \left[
X^2 - (X_{eq})^2 \right] + \frac{B_{\chi}}{m_{\chi}}\left (
\frac{\pi^2 g_*}{30}\right )^{1/2} \frac{A^{1/2} \Phi
T_{RH}}{\sqrt{\Phi + R/A + X \langle E_{\chi}\rangle/T_{RH}}}. %
\label{Boltz3}
\eea%

We now begin to examine the relic densities of the predicted
candidate $\chi \equiv S_1$ of the $B-L$ model using the above set
of Boltzmann equations (\ref{Boltz1}-\ref{Boltz3}) and the
annihilation cross section defined by
Eq.~(\ref{S1sigma}).  We plot the inferred density as a function
of reheat temperature in Fig.\ref{BLabundance1}. Here we fix the
mass of the WDM candidate as 1 keV. This is taken as a reference, as  results are easily extrapolated 
for the mass range relevant
to WDM; since (as in Fig.~1), for the mass range
$\sim 1-100$ keV, the inferred abundances simply scale as
$m_\chi$. Having fixed the mass, we consider the effect of
different values of the $B_\chi$-parameter, which refers to the
branching ratio of $\phi$-decay into $\chi \chi$. In addition, we
include the case with $B_\chi=0$, which is analogous to what was
considered in section III, with standard Fermi coupling. We have
used the average WDM annihilation cross section as defined in
Eq.~(\ref{S1sigma}), with $M_{Z'}=600$ GeV and $g_{_{B-L}}= 0.1$.

\begin{figure}[t]
\begin{center}
\epsfig{file=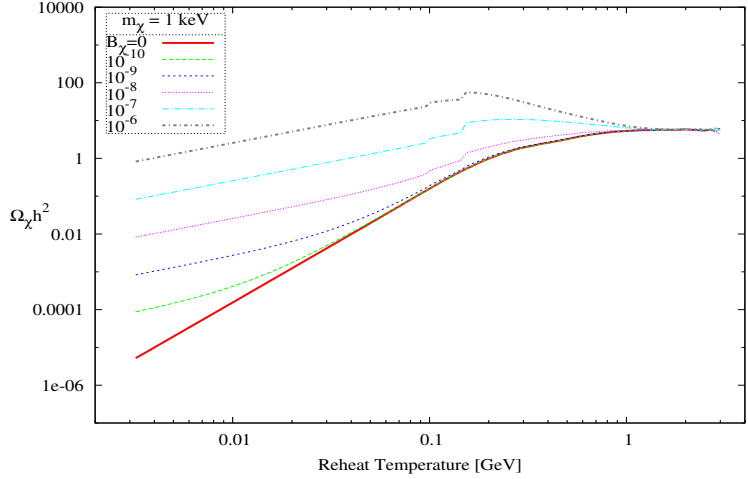,height=6.5cm,width=10.5cm,angle=0}
\caption{Thermal ($B_{\chi}=0)$ and non-thermal relic abundance for $B-L$
light sterile neutrino $S_1$ as function of the reheat
temperature and different branching ratios of decay of the $\phi$
field into WDM particles.}
\label{BLabundance1}%
\end{center}
\end{figure}

First thing to note is that with $B_\chi \lsim 10 ^{-7}$ and low
reheating, $T_{RH} \lsim {\cal O}(0.01)$ GeV, one can easily
account for the observed relic abundance and get $\Omega h^2
\simeq 0.1$. While with $B_\chi =0$, a larger reheat
temperature, $T_{RH} \sim {\cal O}(1)$ GeV is required. 
Since the
annihilation cross section of $S_1$ is  five orders of
magnitudes smaller than the Fermi coupling cross section that has
been considered in Section III, the relic density of $S_1$ WDM is
also about five orders smaller than that used  in Fig.~\ref{Fermi}; as the
abundance, in this case, is proportional to $\langle \sigma v
\rangle$ as can be seen from Eq.~(\ref{ab:red}). This means, as is apparent  from Fig.
\ref{BLabundance1}, that the relic abundance associated with  WDM of
mass $\sim 1$ keV can be consistent with the observational limits
for $T_{RH}$  well above the Big Bang nucleosynthesis
constraint: $T_{RH} \gsim 1$~MeV. Moreover, the whole mass range
 compatible with WDM  can give rise to abundances 
consistent  with empirical constraints on DM density and reheat temperature. This, as opposed to the case of standard
Fermi coupling where the allowable range of relevant masses lay only in the
keV range - rendering them in tension with lower mass limits
inferred from Lyman-$\alpha$ observations - and with reheat
temperatures marginally consistent with Big Bang nucleosynthesis.

When $B_\chi\ne 0$, the behavior of the curves in
Fig.~\ref{BLabundance1} is straightforward to explain in terms of
non-thermal out of equilibrium production - in which case
$\Omega_{\chi} h^2 \propto T_{RH}$ - and non-thermal particle
production while the coupling is strong enough to maintain
chemical equilibrium, in which case $\Omega_{\chi} h^2 \propto
T_{RH}^{-1}$~\cite{Gelmini:2006pw}. When production is dominated
by the non-thermal channel,  emanating from the decaying $\phi$-field,
$\Omega_{\chi} h^2$ is proportional to $ \propto B_\chi$, for a given reheat temperature.
Bearing these scaling relations in mind (and recalling that
$\Omega_\chi h^2 \sim m_\chi$), the minimal case (of $m_\chi =
1$~keV) described here can be used to put constraints on the
branching ratio of $\phi$ field decay into WDM: for a given value
of $m_\chi$ and $T_{RH}$, $B_\chi$ has to be below a certain (rather small)
value to avoid over-production of WDM particles.

\begin{figure}
\begin{center}
\epsfig{file=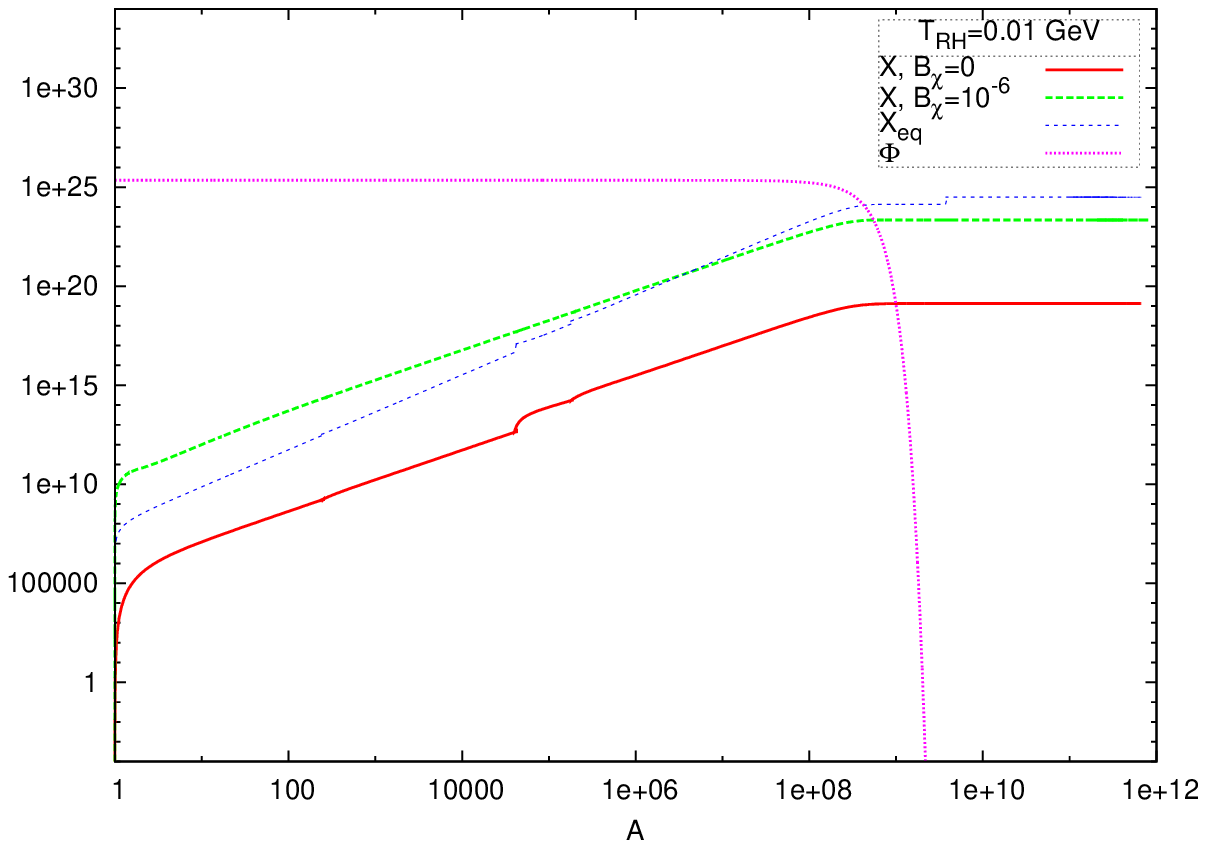,height=6.cm,width=8cm,angle=0}
\epsfig{file=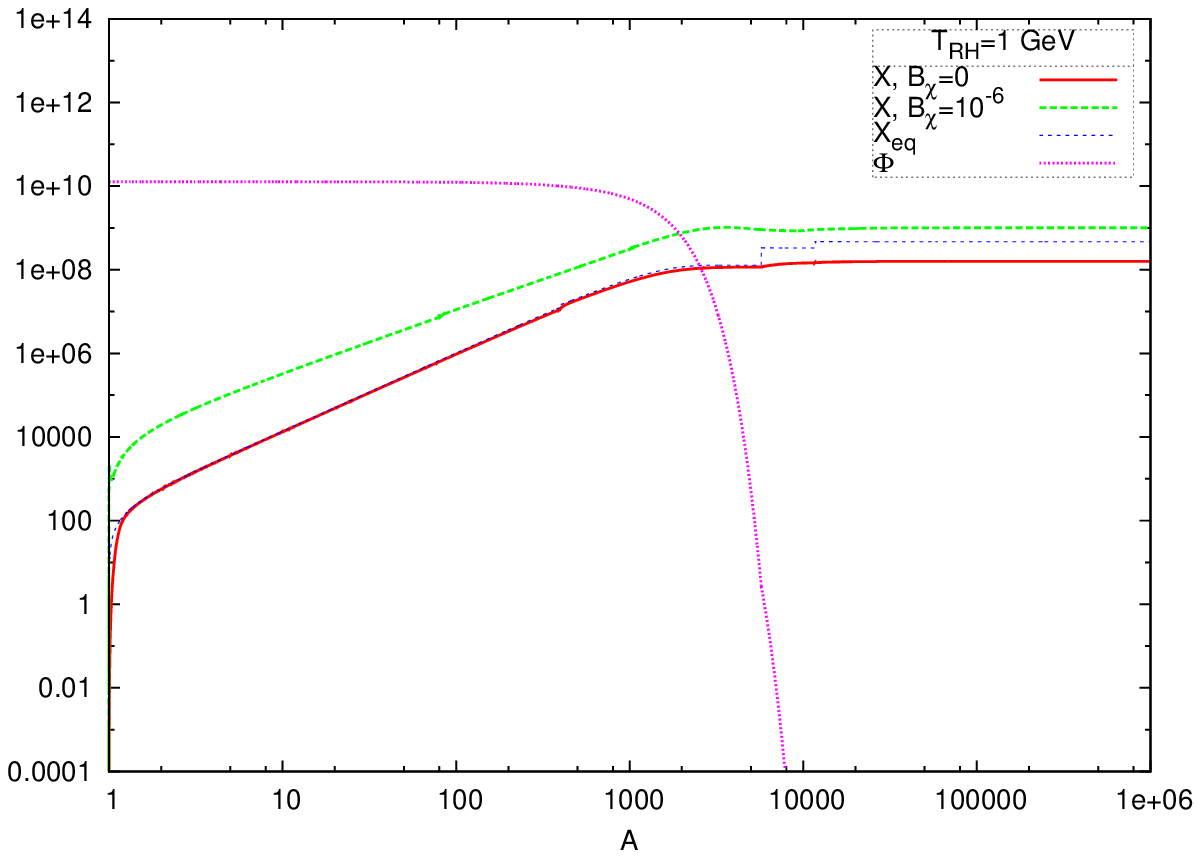,height=6.cm,width=8cm,angle=0}
\caption{Evolution of the comoving abundance of a particle of
mass 1~keV, as compared to the equilibrium abundance. Also shown is the
evolution of the $\phi$ field in terms of $\Phi$.}
\label{A_evol}
\end{center}
\end{figure}

We now take a more detailed look at the dynamics of the particle
production mechanism. As we will see, the relevant processes
crucially depend on the dynamics at reheating, because most of
particle production  occurs just before, or just after reheating
is completed, that is at $T \sim T_{RH}$. In Fig.~\ref{A_evol} we
plot the evolution of the comoving densities $X$ and $X_{eq}$ and
 $\Phi$ (for the decaying scalar field $\phi$)  as
functions of the normalized scale factor $A = a/a_I$ (cf.
Section~\ref{sec:low}). Because $T_{RH} \gg m_{\chi}$, particles
in kinetic equilibrium are relativistic. And since the
relativistic equilibrium density $n_{eq} \sim T^3$, it follows
that for relativistic particles $X_{eq} = n_{eq} a^3$ is constant
during the radiation era (when $T \ll T_{RH}$), since in that case
$T \sim 1/a \sim 1/A$. This explains the constant plateaus in the comoving
densities for large $A$. However, as long as the $\phi$ field is
dominant, $T \sim A^{-3/8}$~\cite{GKR}, with the implication that  $X_{eq} \sim A^{15/8}$,
which is consistent with the initial behavior (for relatively
small $A$) of the equilibrium density, as shown in
Fig.~\ref{A_evol}.

The behavior of $X$  prior to reheating  in the various cases can be largely understood
by considering Eq.~(\ref{Boltz3}), while assuming the $\phi$-field
is dominant. In this case it can be written as %
\be %
\frac{dX}{dA} = - C A^{-5/2} \langle \sigma_\chi v \rangle (X^2 -
X_{eq}^2) + D A^{1/2}, %
\label{ab:red} %
\ee %
where $C = \sqrt{\frac{3}{\Phi_I}} M_{P} T_{RH}$ and $D=
\frac{\pi B_\chi T_{RH}} {m_\chi} \sqrt{\frac{\Phi_I g_*}{30}}$
are constants (barring variations in $g_*$). In the case of
relatively small reheat temperature (say $T_{RH} \sim 0.01$~
GeV), the coupling $C$ is modest and the comoving density $X$
never reaches chemical equilibrium values; and so $X \ll X_{eq}$.
In that case, and if $B_\chi = 0$, Eq.~(\ref{ab:red}) implies that
$\frac{dX}{dA} \sim A^{-5/2} \langle \sigma_{\chi} v \rangle
X_{eq}^2$, which scales as $A^{1/2}$ (since, from Fig.~3, $\langle
\sigma_{\chi} v \rangle \sim T^2 \sim A^{-3/4}$ in the relevant
temperature range).  This in turn implies $X \sim A^{3/2}$, which
agrees with the corresponding curve shown on the left hand panel
of Fig.~\ref{A_evol}. The same scaling is present if $B_\chi$ is
non-zero and large enough; in this case the second term of
Eq.~(\ref{ab:red}) is dominant and $X \sim B_\chi A^{3/2}$; i.e.,
with the same scaling as the previous case but with different
normalization.

For high $T_{\rm RH} \sim 1$~GeV, the coupling $C$ is strong
enough; so that $X$ may quickly reach chemical equilibrium values
when $B_\chi=0$; and so we find that (right hand panel of
Fig.~\ref{A_evol}) $X \rightarrow X_{eq}$ for most of the
evolution (with minor deviations at large $A$, when $X_{eq}$ is
affected by temperature jumps due to variations in $g_*$ while the
decoupled $X$ values remain constant after freeze out)~\footnote{The sharp jumps observed on the plots of Fig. \ref{A_evol} and 6 originate from the sudden changes in the effective number of degrees of freedom, 
$g_*$, in course of our numerical calculations. We note that  $g_*$ can actually be a smooth function of temperature in a more sophisticated analysis \cite{Kolb+T}.}. 
On the other hand if $B_\chi$ is large enough, we have 
$X \gg X_{eq}$ which leads to $\frac{dX}{dA} \approx - C A^{-5/2}
\langle \sigma_\chi v \rangle X^2 + D A^{1/2}$. The term $D
A^{1/2}$ now represents a new `effective equilibrium' value that
$X$ tries to reach (so as to minimize $dX/dA$, as in the standard
equilibrium situation when $X \rightarrow X_{eq}$); if this is
the case one might expect that the $X$ evolution should obey 
\be C
A^{-5/2} \langle \sigma_\chi v \rangle X^2 = D A^{1/2}. %
\ee %
Since, again, $\langle \sigma_\chi v \rangle \sim T^2 \sim
A^{-3/4}$; that means that the first term is proportional to
$A^{-13/4} X^2$, which in turns implies that $X \sim A^{15/8}$, in
this regime. This happens to scale the same way with $A$ as
$X_{eq}$, but again with different normalization. When $A$ is
large enough and $\Phi$ has decayed, $X$ converges to the
equilibrium $X_{eq}$ before decoupling and freezing out.

Due to the shapes of the corresponding curves in Fig.~\ref{A_evol} it is not
obvious when `freeze out' occurs. To get a better picture of the
process we have also reproduced (Fig.~\ref{freeze}) the
logarithmic derivative $d \log X/ d \log A = \frac{A}{X} \frac{dX}{dA}$ as a function of $A$.
If freeze out happens before reheating is complete, 
 $dX/dA$ will tend to zero even though $X_{eq}$  is still varying.
 In the terms of the scaled or normalized measure of the logarithmic
derivative, freeze out can be defined as to occur when this is significantly
smaller than one.

In Fig.~\ref{freeze} we plot the evolution of the logarithmic
derivative $d \log X/ d \log A$ versus $A$ for the cases with
$B_\chi = 0$ (the effect of the decaying field, represented by a
non-zero $B_\chi$, clearly always freezes out when $\Phi
\rightarrow 0$). We also add the evolution of the quantity $x =
m_X/ T$. The inflection point in $d \log X/ d \log A$ evolution
represents the end of the  transition to the
radiation dominated era (and this, as must be the case happens
around $T_{RH}$, as represented by the horizontal lines). In the case
of $T_{\rm RH} = 0.01$ ~GeV, one finds that the freeze out occurs
just around this transition, while for $T_{\rm RH} = 1$ ~GeV, it
occurs significantly later.
In both cases freeze out occurs at
quite small values of $x$, suggesting relativistic decoupling near $T_{RH}$. 

\begin{figure}
\begin{center}
\epsfig{file= 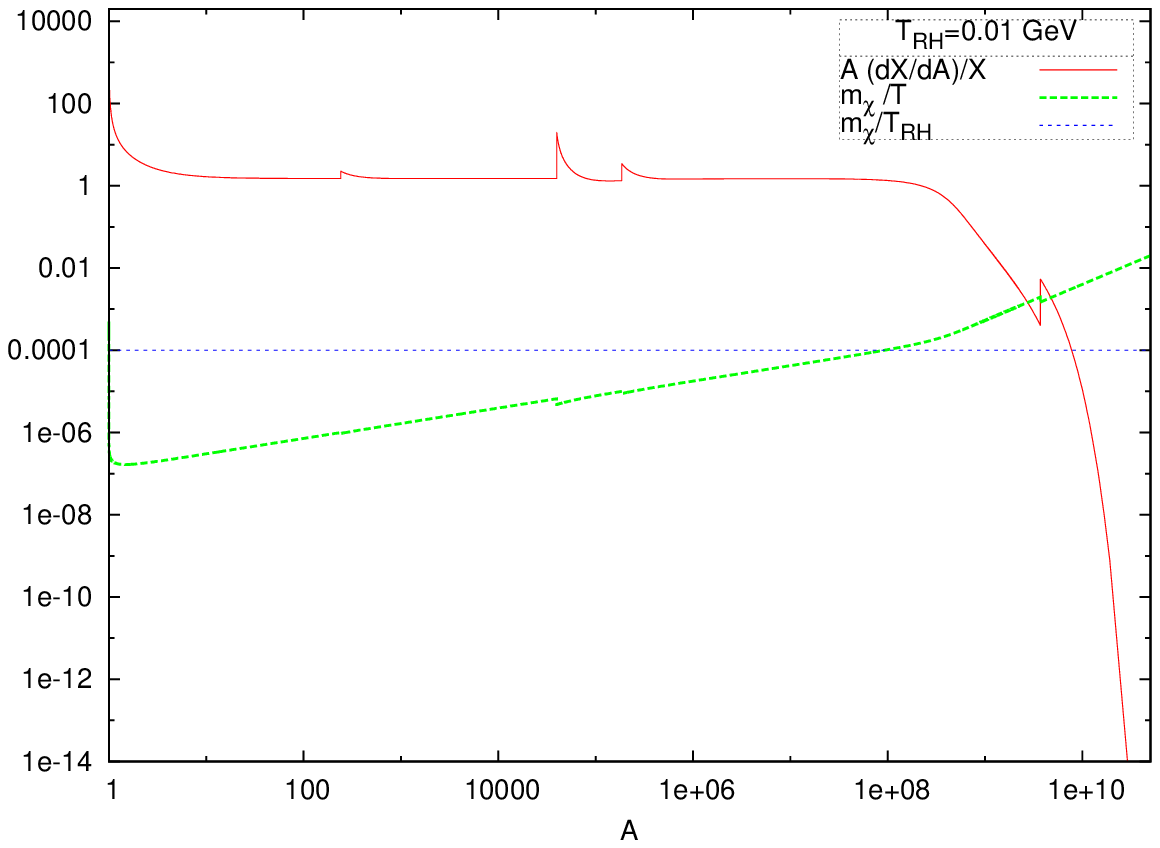,height=6.cm,width=8cm,angle=0}
\epsfig{file= 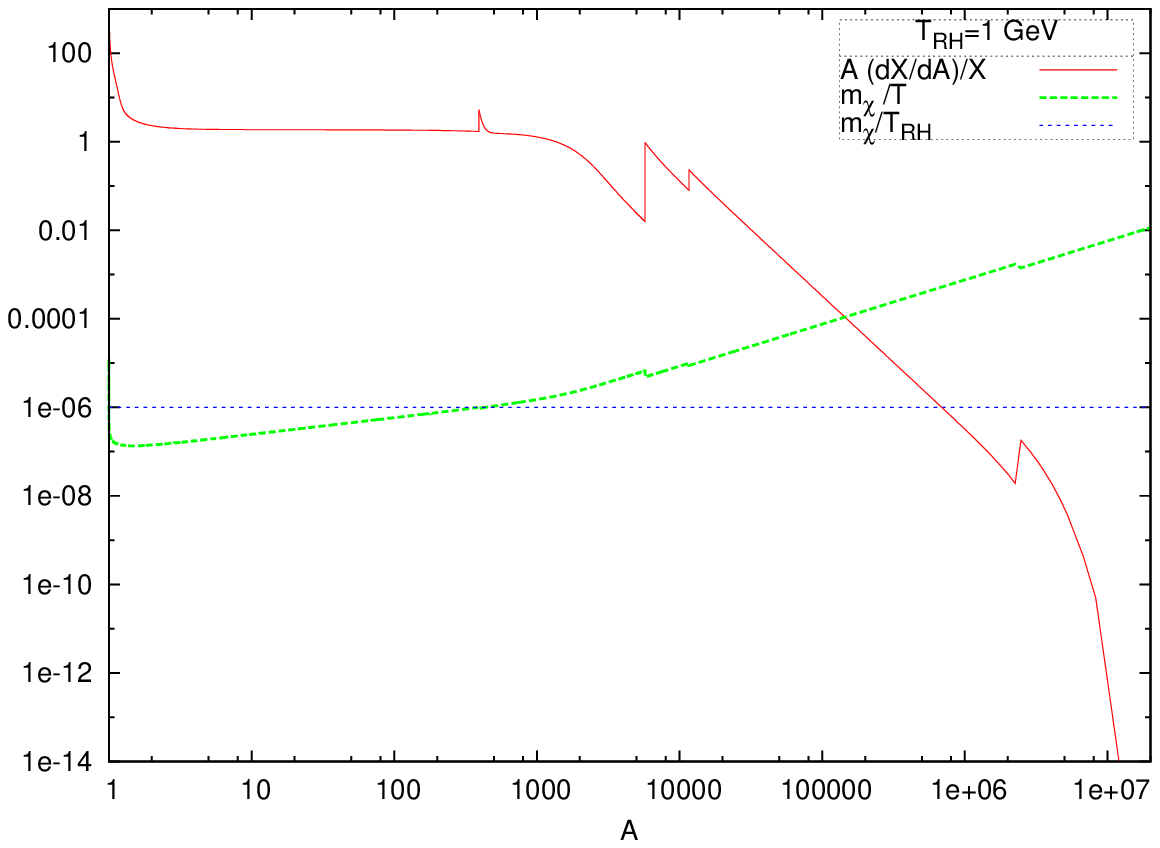
,height=6.cm,width=8cm,angle=0}
\label{Labundance1}%
\caption{Evolution of the logarithmic derivative of the abundance
of a particle of 1~keV, as compared to the equilibrium abundance
for $B_\chi=0$ with $T_{RH}=0.01$ GeV (left panel) and $T_{RH}=1$
GeV (right panel). Also shown is the evolution of the quantity $x
= m_X/T$. } \label{freeze}
\end{center}\end{figure}

We have thus far assumed that kinetic equilibrium is established even 
when the WDM particles are produced non-thermally through $\phi$-field decay; but 
given that our WDM particle interactions are strongly suppressed, this may 
not always be the case. In the absence of kinetic equilibrium, memory of the 
initial conditions of particle production is not lost, and so the WDM streaming 
length will depend on the initial kinetic energy imparted to the DM particles 
at production and on the epoch when this takes place. The streaming length will 
thus depend on the mass of the $\phi$-field and the reheat temperature, in 
addition to its dependence on the mass of $\chi$-particle. 

An estimate of whether kinetic equilibrium is actually established \cite{chen}, 
\cite {kawasaki} can be obtained by applying the usual condition requiring 
that the scattering rate 
of the dark matter with other particles  be larger than the expansion rate, 
$\Gamma^{\rm scatt} > H$; where $H$ is given by Eq.(\ref{H-defn}). 
The scattering rate is related to the scattering cross section ($\sigma_s$) 
and number density of scattering particles ($n$) as $ \Gamma^{\rm scatt} = \sigma_s  n$.  
Our proposed DM particles are expected to scatter off  other particles 
(through the $t$-channel exchange of $Z'$) with an estimate given by 
$\sigma_s \sim g^4_{B-L} \frac{E^2_{\chi}}{M^2_{Z'}}$ (neglecting the mass of the
$\chi$ particle), where $E_{\chi}$ is 
the energy associated with the $\chi$ particles. Considering 
$n \sim T^3$ and replacing $E_{\chi}$ by temperature $T$, 
we can write the condition for kinetic equilibrium at temperature $T = T_{RH}$ as:
\begin{equation}
 T^3_{RH} M_P > \left( \frac{M_{Z'}}{g_{B-L}}\right)^4 .
\end{equation} 
With  $g_{B-L}/M_{Z'} \sim (6 {\rm{TeV}})^{-1}$, this condition is  satisfied for a reheat temperature 
$T_{RH}  > 0.1$ GeV.  And it is sufficient that this condition be satisfied at reheating, following the 
$\phi$-field decay, for the WDM particles to attain kinetic equilibrium. 
Furthermore, one may expect  that kinetic equilibrium is rapidly established as the 
reheat temperature increases above the limit just derived, since the ratio of the 
scattering rate to the expansion rate at reheating increases as $T_{RH}^3$; in this 
case, the streaming length will depend only on the mass of the $\chi$-particle; and 
as mentioned in the introduction, a generous WDM mass range is between 1 and 100 keV. 
However the range of reheat temperature we are interested in covers $T_{RH}$ between 
0.01 GeV to 1 GeV. The steep temperature dependence should also ensure that significantly 
below the above critical temperature one can calculate the free streaming length while  
assuming that decay products of the $\phi$-field do not scatter at all (the intermediate 
case requires detailed calculation of the distribution function and requires a separate 
study). As mentioned above, since kinetic equilibrium is not established, one may 
expect the  streaming length will depend on the mass  of the 
$\phi$-field and the reheat temperature, as well as the mass of the $\chi$ particle --
as indeed is found to be the case \cite{takahashi}. 
For example, using the inequality in Eq.(15) of~\cite{takahashi}, and  
taking  $m_\phi = 1~{\rm TeV}$ and $T_{\rm RH} = 0.01 ~{\rm GeV}$, we can estimate that our $\chi$ particles  
are WDM, in the sense satisfying the relevant Lyman alpha limit, if  
\be
\left(\frac{m_{\chi}}{\rm keV}\right) \gsim 2 \times 10^4  \epsilon,
\ee
where $\epsilon  m_\phi/2$ is the energy imparted into a WDM particle by the decay of a $\phi$ particle. 
That is, for keV mass particles, the streaming length would be  compatible with Lyman-$\alpha$ bounds \cite{Viel:2006kd, VielN}, 
if $\epsilon \sim 10^{-4}$.

Finally we note that the equations ($e.g.$ Eq.~(\ref{eqchi})) we have used to deduce 
the abundances are derived under the assumptions (a) the $\chi$ particles are created 
from and annihilate into particles that constitute a thermal bath, and (b) the 
$\chi$ particles themselves are in kinetic equilibrium. It is in this context
that the thermally averaged cross section arises in those  equations (instead 
of a cross section averaged over a general phase space distribution function). 
One may therefore ask if this 
formulation can still be used when assumption (b) is not  
satisfied, $i.e.$ when $T_{RH} < 0.1$~GeV. To answer this question we note that, when deriving Eq.(\ref{eqchi})  
from the full Boltzmann equations (involving the phase space distribution functions),
assumption (b) is only necessary for deriving the terms involving  $(n_\chi)^2$; in contrast, 
terms with $(n_\chi^{\rm eq})^2$ are derived solely from assumption (a)~\cite{Kolb+T}. 
Now, these  two terms in Eq.(\ref{eqchi})
translate into corresponding ones involving  $X^2$ and $X_{\rm eq}^2$ in Eq.~(\ref{ab:red}). 
Following the discussion subsequent to Eq.(\ref{ab:red}), it is apparent that, for 
low $T_{\rm RH}$  (and small $B_\chi$), our numerical results  can be well understood 
by  assuming $X \simeq 0$, as during relevant evolution $X \ll X_{\rm eq}$, which can be seen 
from Fig.{\ref{A_evol}} (left hand panel), where the difference between $X_{ \rm eq}^2$  and $X^2$ can be inferred to exceed ten orders of magnitude 
(for small $T_{\rm RH}$ and non-negligible $B_\chi$, $\chi$-particle production is dominated by direct 
$\Phi$-field decay, and therefore the manner in which the cross section is averaged over the phase space distribution makes little 
difference). Given this, it seems sufficient that the background bath be a thermal one for our 
abundance calculations to be valid to a good approximation at low $T_{\rm RH}$, since the term associated with  assumption (b) 
should be small~\footnote{Note that if the cross section, averaged over the actual phase space distribution function is 
much larger than the thermally averaged one, then the effect of this term will be to suppress the abundance even further. 
In that case our inferred abundances, which are already small when  $T_{\rm RH} <0.1$~GeV and $B_\chi =0$, can be regarded as upper 
limits, and the conclusions outlined below remain valid. It might also be pertinent to mention here that the assumption 
$\langle E_{\chi} \rangle \simeq \sqrt{{m_{\chi}^2} + 9 T^2}$ also breaks down when $\chi$ is not in kinetic equilibrium. 
However time evolution of radiation is mainly governed by the other terms involved in the Eq.(\ref{eqchi}), 
and hence does not have much impact on the study.}.

\section{Conclusion}
In the standard scenario, with high reheat temperature and
equilibrium freeze out subsequent to reheating, the relic density
of WDM with mass of order ${\cal O}(1)$ keV is unrealistically
higher by several orders of magnitude than the observational
limit. We have shown that, within low reheating scenario, the relic
abundance of dark matter with mass of order keV or smaller or 10 MeV or larger, and
annihilation cross section corresponding to standard Fermi coupling may be
compatible with observations. However, keV WDM is in tension with
Lyman-$\alpha$ and phase space density constraints due to its low
mass and correspondingly large steaming length; the second case, of relatively high mass particles, is not
relevant for solving problems related to the formation of
galaxies, such as the overabundance of low mass DM halos or their
apparent excessive concentrations. Indeed they are already non-relativistic at
reheating and their steaming length will be too small to have any
effect on the density fluctuation power spectrum as compared to
standard CDM. Moreover, in both cases the required reheat
temperatures are so low as to be marginally compatible with Big
Bang nucleosynthesis constraints.

In contrast, due to a much smaller annihilation cross section,
relic densities inferred in the context of the $B-L$ model are
compatible with observations for the whole range of particle
masses relevant to structure formation in the WDM scenario and
with reheat temperatures compatible with Big Bang
nucleosyntheis constraints. Our scenario also allows for the decay
of the $\phi$ field into WDM particles during reheating. The
results sensitively depend on the value of the branching ratio.
Observational bounds on the possible relic abundance can be thus
used to put upper limits on possible values of the branching
ratios.\\ 

\noindent {\bf{Acknowledgments}}:  S.K. would like to acknowledge partial support by 
European Union FP7 ITN INVISIBLES (Marie Curie Actions, PITN-GA-2011-289442). A.S 
acknowledges the hospitality of CTP, Zewail City for Science and Technology, 
Cairo during a visit when this work was initiated. We thank  
the referee for careful reading of the manuscript and comments that 
helped improve the paper.


\end{document}